\documentclass{sf2a-conf2012}
\usepackage{graphicx}
\usepackage{hyperref}
\usepackage[]{natbib}  
\usepackage[cyr]{aeguill}
\usepackage{epstopdf}
\usepackage{times}
\usepackage{amssymb,amsmath}

\def\BibTeX{{\rm B\kern-.05em{\sc i\kern-.025em b}\kern-.08em
    T\kern-.1667em\lower.7ex\hbox{E}\kern-.125emX}}
\bibpunct{(}{)}{;}{a}{}{,}  

\begin{document}

\TitreGlobal{SF2A 2012}


\title{The accretion disc, jets and environment of the intermediate mass black hole candidate ESO 243-49 HLX-1}

\runningtitle{The IMBH ESO 243-49 HLX-1}

\address{Universit\'e de Toulouse; UPS-OMP; IRAP, Toulouse, France}
\author{N.A. Webb$^{1,}$}\address{CNRS; IRAP; 9 avenue du Colonel Roche, BP 44346, F-31028 Toulouse Cedex 4, France}
\author{D. Barret$^{1,2}$}

\author{V. Braito}\address{Department of Physics and Astronomy, University of Leicester, University Road, Leicester LE1 7RH, UK}
\author{S. Corbel}\address{Laboratoire Astrophysique des Interactions Multi-echelles (UMR 7158), CEA/DSM-CNRS-Universit\'e Paris Diderot, CEA Saclay, F-91191 Gif sur Yvette, France}
\author{D. Cseh$^{4}$}
\author{S. A. Farrell}\address{Sydney Institute for Astronomy, School of Physics, The University of  Sydney, NSW 2006, Australia}
\author{R.P. Fender}\address{School of Physics and Astronomy, University of Southampton, Highfield, Southampton, SO17 1BJ, UK}
\author{N. Gehrels}\address{Astroparticle Physics Laboratory, NASA/Goddard Space Flight Center, Greenbelt, MD 20771, USA}

\author{O. Godet$^{1,2}$}

\author{I. Heywood}\address{University of Oxford, Department of Physics, Keble Road, Oxford OX1 3RH, UK}
\author{T. Kawaguchi}\address{Center for Computational Sciences, University of Tsukuba, 1-1-1 Tennodai, Tsukuba, Ibaraki 305-8577, Japan}
\address{Institut d'Astrophysique de Paris, UMR 7095 CNRS, UPMC Université Paris 06, 98bis Boulevard Arago, 75014 Paris, France }
\author{J.-P. Lasota$^{10,}$}\address{Astronomical Observatory, Jagiellonian University, ul. Orla 171, 30-244 Krakow, Poland}
\author{E. Lenc}\address{Australia Telescope National Facility, CSIRO Astronomy and Space Science, PO Box 76, Epping NSW 1710, Australia}
\author{D. Lin$^{1,2}$}
\author{B. Plazolles$^{1,2}$}
\author{M. Servillat$^{4}$}




\setcounter{page}{237}


\maketitle


\begin{abstract}
The Ultra Luminous X-ray (ULX) source HLX-1 in the galaxy ESO 243-49 has an observed maximum unabsorbed X-ray luminosity of 1.3 $\times$ 10$^{42}$ erg/s (0.2-10.0 keV). From the conservative assumption that this value exceeds the Eddington limit by at most a factor of 10, the minimum mass is then 500 M$_\odot$. The X-ray luminosity varies by a factor of 40 with an apparent recurrence timescale of approximately one year. This X-ray
variability is associated with spectral state transitions similar to those seen in black hole X-ray binaries. Here we discuss our recent modelling of all the X-ray data for HLX-1 and show that it supports the idea that this ULX is powered by sub- and near Eddington accretion onto an intermediate mass black hole. We also present evidence for transient radio emission which is consistent with a discrete jet ejection event as well as comment on the nature of the environment around HLX-1 in light of recent Hubble Space Telescope photometry.

\end{abstract}



\section{Introduction}
 Two varieties of black holes (BHs) have been observed to date: stellar
mass ($\sim$3-20 M$_{\odot}$) BHs and supermassive ($\sim$
10$^{5-10}$ M$_{\odot}$) BHs present in the cores of most large
galaxies. It is believed that stellar mass BHs are formed from
the collapse of massive stars \citep[e.g.][]{frye03}, but
it is not yet clear how supermassive ones are formed. One
model proposes that they are formed from the mergers of smaller mass
($\sim$10$^{2-5}$ M$_{\odot}$) BHs, the so-called
intermediate mass black holes  \citep[IMBHs, e.g.][]{mada01}. Another model proposes super-Eddington accretion onto smaller mass
BHs, to form supermassive black holes, which would again imply the existence of IMBH \citep{kawa04}. But if either of these scenarios are correct, why have IMBH not been observed?  How IMBHs form and where they reside is also a
subject of intense debate \citep{mill04}. They may be formed through accretion and reside in the centres of old dense stellar clusters (globular clusters) or directly in young star forming regions \citep{mill04}. IMBHs constitute the missing BH link.  They are also of interest to a wide variety of fundamental physics and astrophysical topics. It is thought that they may play a role in the stability of globular clusters \citep[e.g.][]{hut92}, that they could be strong sources of gravitational waves if they are in elliptical binary systems with other compact objects \citep{mill04}, that dark matter may be comparatively easy to detect around them  \citep[e.g.][]{forn08}, that gas accretion onto IMBHs should have contributed significantly to the UV background \citep{kawa03} and that IMBHs may have participated in the cosmological ionisation \citep{mada04}.

Ultra-luminous X-ray sources (ULXs) are non-nuclear extragalactic objects with bolometric luminosities
$>$10$^{39}$ erg s$^{-1}$ \citep[e.g.][]{robe07}. Luminosities up to $\sim$10$^{41}$
erg s$^{-1}$ can be plausibly explained through beaming effects \citep{king08,free06} and/or hyper-accretion onto stellar mass BHs
\citep{king08,kawa03,bege02}. A rare class of ULX -- the hyper-luminous
X-ray sources -- have X-ray luminosities $>$10$^{41}$ erg s$^{-1}$
and require increasingly complicated and unlikely scenarios to explain them without
invoking the presence of an IMBH. We discovered a hyper luminous X-ray source (Farrell et al. 2009), dubbed HLX-1, consistent with being in the edge-on early-type galaxy ESO 243-49 at 95 Mpc, thanks to our FORS2 spectroscopy of the faint optical counterpart to HLX-1 \citep{wier10, sori10}.  Using the maximum unabsorbed X-ray luminosity of $\sim$10$^{42}$ erg
s$^{-1}$ (0.2-10.0 keV, Farrell et al. 2009) and the
conservative assumption that it exceeds the Eddington limit by
at most a factor of 10 \citep{bege02}, a lower mass limit of 500
M$_\odot$ was derived for the BH \citep{farr09}.

Our regular monitoring of HLX-1 with {\em Swift} \citep{gehr04,burr05} has revealed significant flux changes , see Fig.~1, in conjunction with simultaneous spectral changes in the same way as Galactic BH X-ray binaries \citep{gode09}, thus strengthening the case for an accreting
BH in HLX-1.  From the four well sampled outbursts along with two prior to these, it has become evident that HLX-1's X-ray variability follows a fairly distinct pattern over approximately 1 year \citep{laso11}.

\section{X-ray data: Modelling the disc emission}
Over the last four years we have carried out a monitoring campaign of HLX-1 with the {\em Swift} X-ray Telescope. During this time HLX-1 has shown four fast rise and exponential decay (FRED) type X-ray outbursts, with increases in the count rate of a factor 40. We have also obtained two {\em XMM-Newton}
and two {\em Chandra} dedicated pointings that were triggered at the lowest and highest luminosities. From simple spectral fitting, the unabsorbed luminosities range from 1.9 $\times$ 10$^{40}$ to 1.3 $\times$ 10$^{42}$ erg s$^{-1}$. Using these data we confirm the proposed spectral state transitions from HLX-1. At high luminosities, the X-ray spectrum shows a thermal state dominated by a disc component with temperatures $\le$0.26 keV, and at low luminosities the
spectrum is dominated by a hard power law with 1.4 $\le \Gamma \le$ 2.1, consistent with
a hard state. The source was also observed in a state consistent with the steep power law state, with $\Gamma$=3.3$\pm$0.2. In the thermal state, the luminosity of the disc component appears to
scale with the fourth power of the inner disc temperature which supports the presence of an optically
thick, geometrically thin accretion disc. The low fractional variability (rms of 9$\pm$9\%) in this state
also suggests the presence of a dominant disc. The spectral changes and long-term variability can not be explained by variations of the beaming and are not consistent with the source being
in a super-Eddington accretion state as is proposed for most ULX sources with lower luminosities.  HLX-1 is therefore an unusual ULX as it is similar to Galactic black hole binaries, which
have, for the most part, non-beamed and sub-Eddington emission. However, HLX-1 differs from them as it has a luminosity three orders of magnitude higher. Comparing HLX-1 to the Galactic black hole binaries, from Eddington scaling we determine a lower limit on the mass of the black hole of $>$9000 M$_\odot$. The
relatively low disc temperature in the thermal state also suggests the presence of an IMBH of few 10$^3$ M$_\odot$ \citep{serv11}.

\begin{figure}[ht!]
 \centering
 \includegraphics[width=1.0\textwidth,clip]{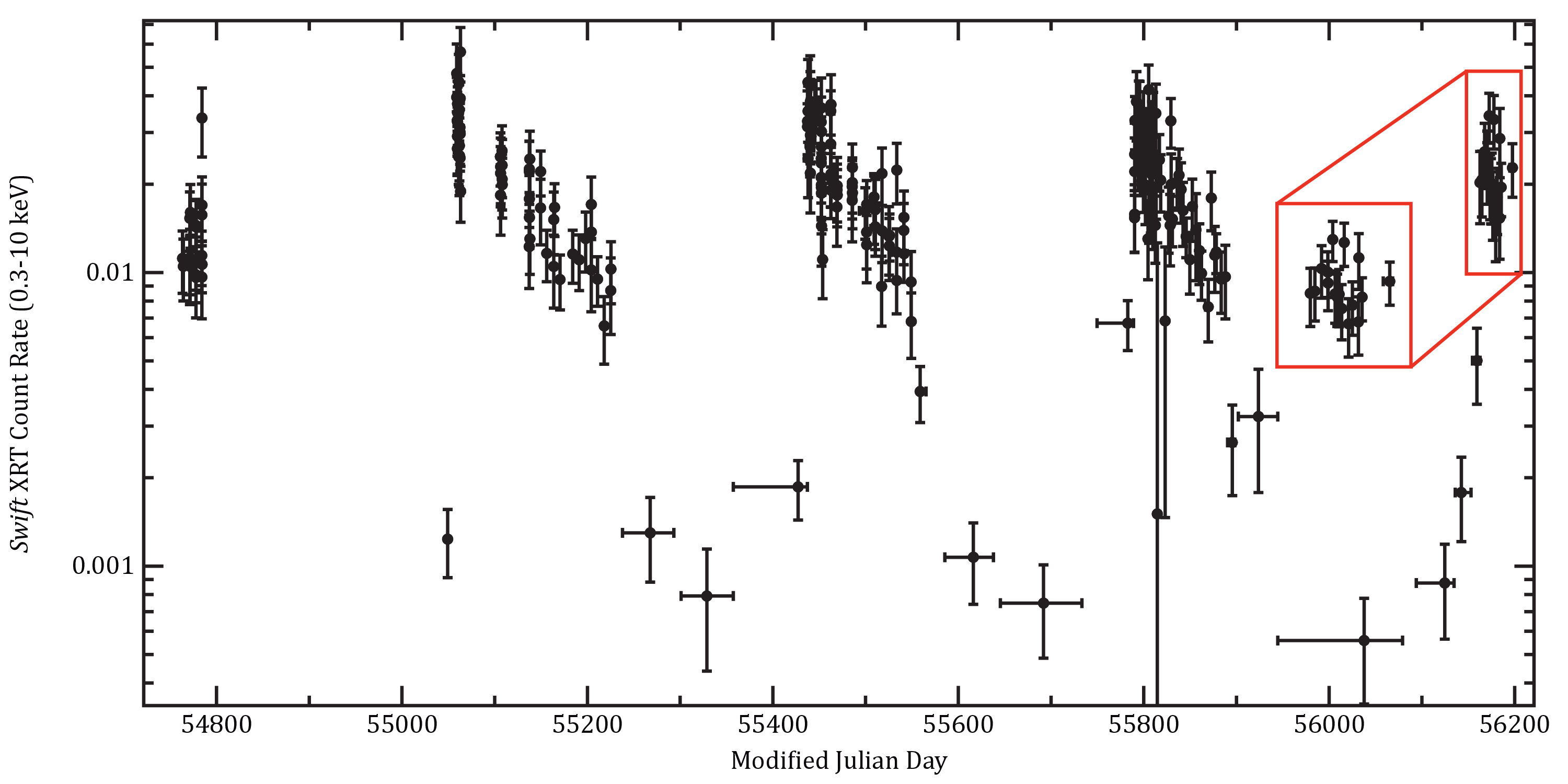}      
  \caption{Longterm Swift lightcurve.  Four X-ray state transitions from the low/hard state (count rate $\lesssim$ 0.002, 0.3-10.0 keV) to the high/soft state (0.01 $\lesssim$ count rate $\lesssim$ 0.05, 0.3-10.0 keV) can be seen.}
  \label{author1:fig1}
\end{figure}

We then fitted more complex models to similar multi-epoch data collected with {\em Swift}, {\em XMM-Newton} and {\em Chandra}. We used a disc model \citep{kawa03} for a wide range of sub- and super-Eddington
accretion rates assuming a non-spinning black hole and a face-on disc (i = 0$^\circ$). Thanks to the wide range of the accretion rates of
this model, one does not need to assume a priori whether
the source is in sub- or super-Eddington phases, unlike
with other disc models. Using this model implies that the black hole in HLX-1 is in the intermediate mass range ($\sim$2$\times$10$^4$ M$_\odot$) and the accretion flow is in the sub-Eddington regime. The disc radiation efficiency is
 $\eta$ = 0.11 $\pm$ 0.03. We also confirm that the source does follow the L$_X$$\propto$T$^4$ relation for this mass
estimate. At the outburst peaks, the source radiates near the Eddington limit. The accretion rate near the Eddington limit
then stays constant around 4$\times$10$^{-4}$ M$_\odot$ yr $^{-1}$ for several days and then decreases exponentially, with a brutal decrease at the end of the outburst that may indicate that the accretion regime changes to an Advection Dominated Accretion Flow (ADAF). {\em Plateaus} in the accretion rate could be evidence that enhanced mass transfer rate is the
driving outburst mechanism in HLX-1 \citep{gode12}. We also obtained good fits to disc-dominated observations of the source with BHSPEC, a fully relativistic black hole accretion disc spectral model \citep{davi11}. Due to degeneracies in the model arising from the lack of independent constraints on inclination and black hole spin, there is a factor of 100 uncertainty in the best-fit black hole mass M. Nevertheless, spectral fitting of {\em XMM-Newton} observations provides robust lower and upper limits with 3000 M$_\odot$ $\lesssim$ M $\lesssim$ 3 $\times$ 10$^5$ M$_\odot$, at 90\% confidence, again placing HLX-1 firmly in the intermediate-mass regime. The upper bound on M is sensitive to the maximum allowed inclination i, and is reduced to M $\lesssim$ 1 $\times$ 10$^5$ M$_\odot$ if the inclination is taken to be below 75$^\circ$. 

\section{Radio emission: Detecting the jets}

We observed HLX-1 with the  Australia Telescope Compact Array (ATCA) in the 750 m configuration on
13 Sep. 2010, when regular X-ray monitoring of HLX-1 with the {\em Swift} satellite  showed that HLX-1 had just undergone a transition from the low/hard X-ray state to the high/soft X-ray state.  The transition occurs for HLX-1 when the count rate increases by more than a factor 10 in just a few days (Fig 1) \citep{gode12,serv11}.  Galactic
BH binaries are known to emit radio flares
around the transition from the low/hard to the high/soft state,
\citep[e.g.][]{fend09,corb04}.  These are associated with 
ejection events, where, for example, the jet is expelled which can lead to radio flaring when the higher velocity ejecta may collide with the lower-velocity material produced by the steady jet.  As well as detecting radio emission from the galaxy nucleus, we
detected a radio point source at Right Ascension (RA) = 01$^h$10$^m$28.28$^s$ and declination (dec.) = -46$^\circ$04'22.3'', coincident with the {\em Chandra} X-ray position of HLX-1 \citep{webb10}.  Combining the 5 GHz and 9 GHz data gives a detection of
50~$\mu$Jy/beam, a 1 $\sigma$ noise level of 11~$\mu$Jy, thus a
4.5 $\sigma$ detection at the position of HLX-1, at a time when such emission can be expected \citep[Fig 2, left, Table~1 \&][]{webb12}.  

The radio flares in Galactic BH binaries are typically a
factor 10-100 (and even more) brighter than the
non-flaring radio emission \citep{koer05} and generally
last one to several days, e.g. XTE J1859+226 \citep{broc02}.
Once the high/soft state has been achieved, the core jet is suppressed \citep[e.g.][]{fend09}.  To determine whether  the
radio emission that we detected was transient and thus associated with
a radio flare, we made another observation with the ATCA in the 6 km configuration on 3 Dec.
2010, when HLX-1 was declining from the high/soft state and when no
flaring is expected.  This observation again showed emission from the
nucleus of the galaxy, consistent with that of the previous radio
observation, but revealed no source at the position of HLX-1.   The
3~$\sigma$ non-detection for the combined  5 GHz and 9 GHz data is
36~$\mu$Jy/beam (Fig 2, right \& Table~1). These observations suggest
that the source is variable.  

\begin{figure}[ht!]
 \centering
\hspace*{-0.3cm} \includegraphics[width=0.34\textwidth,angle=-90,clip]{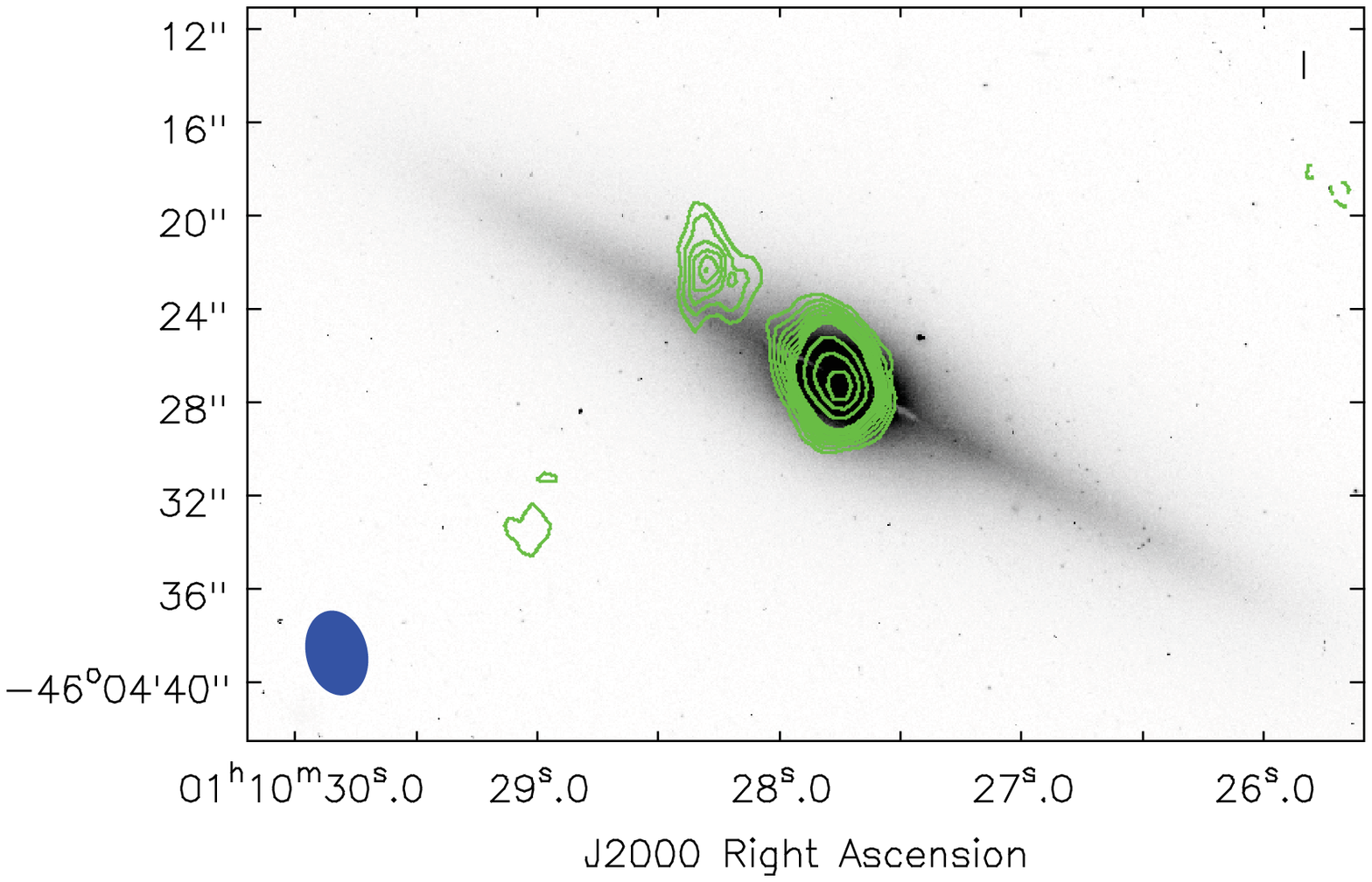}%
 \includegraphics[width=0.34\textwidth,angle=-90,clip]{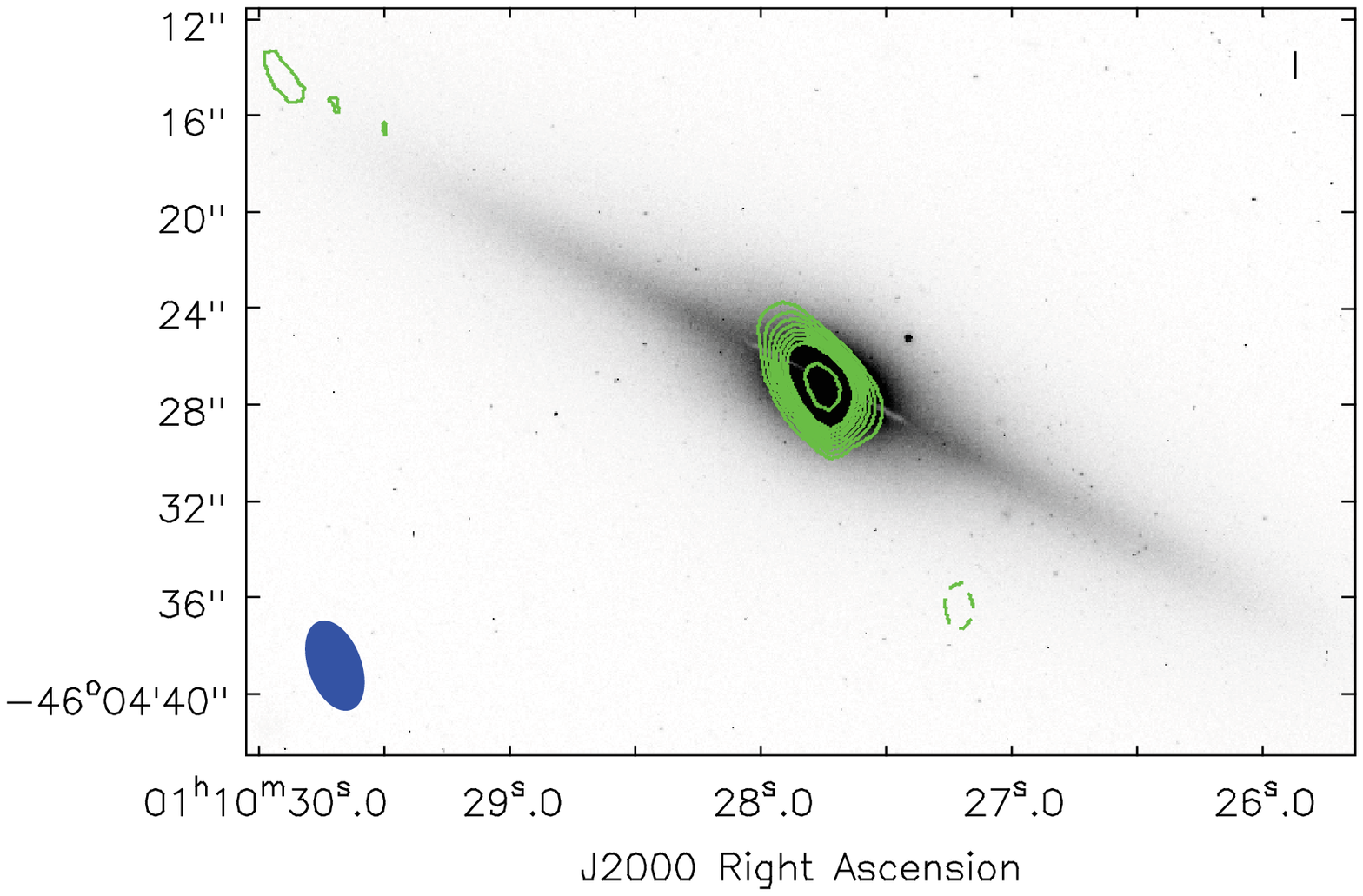}      
  \caption{{\bf Left:} Left: 5 and 9 GHz combined radio observations (contours: -3, 3, 4, 5, 6, 7, 8, 9, 10, 15, 20, 25 times the 1 $\sigma$ rms noise level (5.6 $\mu$Jy/beam)) using radio data taken on the 13th September 2010, 31st August 2011, 3rd and 4th September 2011 with the ATCA and superimposed on an I-band Hubble Space telescope image of ESO 243-49 (inverted colour map). The beam size is shown in the bottom left hand corner.  The galaxy, ESO 243-49, is clearly detected in radio. An 8 $\sigma$ point source falls at RA = 01$^h$10$^m$28.28$^s$ and declination = -46$^\circ$04'22.3'' (1 $\sigma$ error on the position of RA=0.43'' and dec.=0.67''), well within the 0.3'' Chandra error circle of HLX-1. {\bf Right:} 5 and 9 GHz combined radio observations (contours: -3, 3, 4, 5, 6, 7, 8, 9, 10, 15, 20, 25 times the 1 $\sigma$ rms noise level (7.0 $\mu$Jy/beam)) made from the 3rd December 2010, 25th August 2011 and 1st September 2011 ATCA observations and superimposed on the same I-band Hubble Space telescope image of ESO 243-49.  The galaxy ESO 243-49 is again clearly detected, but no source is found within the Chandra error circle.  Again the beam size is shown in the bottom left hand corner. }
  \label{author1:fig2}
\end{figure}

To confirm the variability, we re-observed HLX-1 when it had just undergone another transition from the low/hard X-ray state to the high/soft X-ray state in August 2011 (Fig 1). All five of the 2011 observations (Table 1) were made in a similar configuration to the December 2010 observation.  We observed three non-contiguous detections ($\geq$ 4 $\sigma$) and two non-contiguous non-detections of the source (Table~1).  This indicates that two flares were detected during this period.

To determine if the source was indeed variable, we fitted each observation using a point source, using the point spread function. We used the position of HLX-1 when the source was not detected.  This allowed us to estimate the flux and the associated errors (Table 1) even for a non-detection. We tested whether the data could be fitted with a constant, namely the mean of the data.  We compared these data to the mean flux value using a chi-squared test.   We found a reduced chi-square ($\chi^{\scriptscriptstyle 2}_{\scriptscriptstyle \nu}$) value of 2.5 (5 degrees of freedom) which is much greater than unity, demonstrating that a constant is a poor fit to  the data and supporting the variable nature. Further, combining all of the detections (5 and 9 GHz), the source is observed at 45~$\mu$Jy/beam, with a 1~$\sigma$ noise level of 5.5~$\mu$Jy, which shows a confident detection at the 8~$\sigma$ level.  Combining, in a similar fashion, the data in which no radio emission was detected, we obtained a  3~$\sigma$ upper limit in the combined 5+9 GHz data of 21~$\mu$Jy/beam (Fig 2).  The variability rules out emission from a nebula.   The observed variable radio emission is then again consistent with a transient jet ejection event \citep{webb12}.\\

\hspace*{-0.5cm}
\begin{minipage}{6.9cm}
{\bf Table 1:} The 7 radio observations organised by date and showing the {\em Swift} X-ray unabsorbed flux (0.5$-$10.0 keV) $\times$ 10$^{-13}$
erg cm$^{-2}$ s$^{-1}$ (and the 90\% confidence errors) along with the combined 5 and 9 GHz peak brightness radio flux (with the associated 1 $\sigma$ noise level) or the 3~$\sigma$ upper limit for the non-detections.  The final column gives the radio flux from fitting a point source (using the point spread function) (and the associated 1 $\sigma$ noise level).  
\end{minipage}
\begin{minipage}{8cm}
\hspace*{0.3cm}
\begin{tabular}{cccc}
\hline
Observation & X-ray & 5+9 GHz peak & 5+9 GHz flux \\
date & flux & flux ($\mu$Jy/beam) &  density ($\mu$Jy)\\
\hline
13 Sep. 2010 & 4.57($\pm^{0.68}_{0.50}$) & 50 (11)  & 42 (10)\\
3 Dec. 2010 & 2.40($\pm^{0.60}_{0.50}$) & $<$36 & 11 (20) \\
25 Aug. 2011 & 4.57($\pm$0.30) & $<$30 & 14.5 (7)\\
31 Aug. 2011 & 4.57($\pm$0.30)&  51 (10) & 63 (18)\\
1 Sep. 2011 & 4.57($\pm$0.30)& $<$31 & 25 (10.5) \\
3 Sep. 2011 & 4.57($\pm$0.30)&  45 (10.5) & 43 (10)\\
4 Sep. 2011 & 4.57($\pm$0.30) & 30 (7.5) & 27 (7.5)\\
\hline
\end{tabular}

\medskip

\end{minipage}

\medskip

It has been shown that observations of super massive BHs and stellar mass BHs support the scale invariance of jets \citep{merl03,koer06}.  This was done by comparing X-ray and radio measurements, tracers of
mass accretion rate and kinetic output respectively, with the BH
mass to form a ``fundamental plane of black hole activity''. Under the hypothesis that HLX-1 is indeed an intermediate mass black hole, we can test the proposed relation.  We take what is generally considered to be the maximum mass of intermediate mass black holes, $\sim$1$\times$10$^{5}$~M$_\odot$ \citep{mill04} and the X-ray luminosity, 5.43$\times$10$^{41}$ erg s$^{-1}$ (0.5-10.0 keV), determined from {\em Swift} X-ray telescope observations made at the same time as our radio detection.  Continuum (non flaring) radio emission could then be estimated with  the aforementioned relationship \citep{koer06},
which is based on a sample that includes BHs in all different X-ray
states.  This relation implies a continuum radio emission at the $\sim$20$\mu$Jy level.  This is slightly lower than the 3~$\sigma$ non-flaring upper limit, suggesting that the mass of the BH is likely to be less than $\sim$1$\times$10$^{5}$~M$_\odot$ \citep{webb12}.

Radio flares are seen to occur in Galactic black hole binaries when the X-ray luminosity is 10$-$100 per cent of the Eddington luminosity \citep{fend04}.   HLX-1 has already shown similar behaviour to the  Galactic black hole binaries. Therefore assuming that the radio flares that we observed also occur when the X-ray luminosity is 10$-$100 per cent of the Eddington luminosity indicates a black hole mass between $\sim$9.2 $\times$10$^{3}$~M$_\odot$ and $\sim$9.2 $\times$10$^{4}$~M$_\odot$, commensurate with the mass estimate above and those of \citep{davi11,serv11,gode12} and confirming the intermediate mass black hole status \citep{webb12}.

\section{Multiwavelength data: Identifying the host population}

Fitting the spectral energy distribution from the near infra-red (Hubble Space Telescope) to X-ray ({\em Swift} X-ray Telescope) wavelengths, we showed that the broadband spectrum is not consistent with simple and irradiated disc models, but is well described by a model comprised of an irradiated accretion disc plus a $\sim$10$^6$ M$_\odot$ stellar population. The age of the population cannot be uniquely constrained, with both young ($\sim$13 Myr) and old ($\sim$13 Gyr) stellar populations allowed. However, the old solution requires excessive disc reprocessing and an extremely small disc, so we favor the young solution ($\sim$13 Myr). In addition, the presence of dust lanes and the lack of any nuclear activity from X-ray observations of the host galaxy suggest that a gas-rich minor merger may have taken place less than $\sim$200 Myr ago. A merger event between a dwarf galaxy and ESO 243-49 could explain the presence of the IMBH (remnant of the stripped dwarf galaxy) and the young stellar population (whose formation was triggered by the merger) \citep{farr12}. \cite{sori12} also made ground based VLT U, B, V, R and I band observations of HLX-1 when the X-ray luminosity had dropped to half the peak luminosity.  They found that the optical magnitudes had also dropped by $\sim$1 magnitude. They modelled the Comptonized, irradiated X-ray spectrum of the disc, and found
that the optical luminosity and colours in the 2010 November data were consistent with
emission from the irradiated disc but state that they strongly
rule out the presence of a young superstar cluster, which would be too bright. However, HLX-1 is contaminated by the diffuse galaxy emission in these ground based images and \cite{sori12} do not fit the X-ray and optical
data simultaneously, as in \citep{farr12}, which could cause the differing results.  New {\em HST} and {\em XMM-Newton} data is expected to resolve this issue (Farrell et al. in prep.).

\bibliographystyle{aa}  
\bibliography{webb} 

\begin{thebibliography}{33}
\expandafter\ifx\csname natexlab\endcsname\relax\def\natexlab#1{#1}\fi

\bibitem[{{Begelman}(2002)}]{bege02}
{Begelman}, M.~C. 2002, Astrophys. J., 568, L97

\bibitem[{{Brocksopp} {et~al.}(2002){Brocksopp}, {Fender}, {McCollough},
  {Pooley}, {Rupen}, {Hjellming}, {de la Force}, {Spencer}, {Muxlow},
  {Garrington}, \& {Trushkin}}]{broc02}
{Brocksopp}, C., {Fender}, R.~P., {McCollough}, M., {et~al.} 2002, Mon. Not. R.
  Astron. Soc., 331, 765

\bibitem[{{Burrows} {et~al.}(2005){Burrows}, {Hill}, {Nousek}, {Kennea},
  {Wells}, {Osborne}, {Abbey}, {Beardmore}, {Mukerjee}, {Short}, {Chincarini},
  {Campana}, {Citterio}, {Moretti}, {Pagani}, {Tagliaferri}, {Giommi},
  {Capalbi}, {Tamburelli}, {Angelini}, {Cusumano}, {Br{\"a}uninger}, {Burkert},
  \& {Hartner}}]{burr05}
{Burrows}, D.~N., {Hill}, J.~E., {Nousek}, J.~A., {et~al.} 2005, Space Sci.
  Rev., 120, 165

\bibitem[{{Corbel} {et~al.}(2004){Corbel}, {Fender}, {Tomsick}, {Tzioumis}, \&
  {Tingay}}]{corb04}
{Corbel}, S., {Fender}, R.~P., {Tomsick}, J.~A., {Tzioumis}, A.~K., \&
  {Tingay}, S. 2004, Astrophys. J., 617, 1272

\bibitem[{{Davis} {et~al.}(2011){Davis}, {Narayan}, {Zhu}, {Barret}, {Farrell},
  {Godet}, {Servillat}, \& {Webb}}]{davi11}
{Davis}, S.~W., {Narayan}, R., {Zhu}, Y., {et~al.} 2011, Astrophys. J., 734,
  111

\bibitem[{{Farrell} {et~al.}(2012){Farrell}, {Servillat}, {Pforr}, {Maccarone},
  {Knigge}, {Godet}, {Maraston}, {Webb}, {Barret}, {Gosling}, {Belmont}, \&
  {Wiersema}}]{farr12}
{Farrell}, S.~A., {Servillat}, M., {Pforr}, J., {et~al.} 2012, \apjl, 747, L13

\bibitem[{{Farrell} {et~al.}(2009){Farrell}, {Webb}, {Barret}, {Godet}, \&
  {Rodrigues}}]{farr09}
{Farrell}, S.~A., {Webb}, N.~A., {Barret}, D., {Godet}, O., \& {Rodrigues},
  J.~M. 2009, Nature, 460, 73

\bibitem[{{Fender} {et~al.}(2004){Fender}, {Belloni}, \& {Gallo}}]{fend04}
{Fender}, R.~P., {Belloni}, T.~M., \& {Gallo}, E. 2004, Mon. Not. R. Astron.
  Soc., 355, 1105

\bibitem[{{Fender} {et~al.}(2009){Fender}, {Homan}, \& {Belloni}}]{fend09}
{Fender}, R.~P., {Homan}, J., \& {Belloni}, T.~M. 2009, Mon. Not. R. Astron.
  Soc., 396, 1370

\bibitem[{{Fornasa} \& {Bertone}(2008)}]{forn08}
{Fornasa}, M. \& {Bertone}, G. 2008, International Journal of Modern Physics D,
  17, 1125

\bibitem[{{Freeland} {et~al.}(2006){Freeland}, {Kuncic}, {Soria}, \&
  {Bicknell}}]{free06}
{Freeland}, M., {Kuncic}, Z., {Soria}, R., \& {Bicknell}, G.~V. 2006, Mon. Not.
  R. Astron. Soc., 372, 630

\bibitem[{{Fryer}(2003)}]{frye03}
{Fryer}, C.~L. 2003, Classical and Quantum Gravity, 20, 73

\bibitem[{{Gehrels} {et~al.}(2004){Gehrels}, {Chincarini}, {Giommi}, {Mason},
  {Nousek}, {Wells}, {White}, {Barthelmy}, {Burrows}, {Cominsky}, {Hurley},
  {Marshall}, {M{\'e}sz{\'a}ros}, {Roming}, {Angelini}, {Barbier}, {Belloni},
  {Campana}, {Caraveo}, {Chester}, {Citterio}, {Cline}, {Cropper}, {Cummings},
  {Dean}, {Feigelson}, {Fenimore}, {Frail}, {Fruchter}, {Garmire}, {Gendreau},
  {Ghisellini}, {Greiner}, {Hill}, {Hunsberger}, {Krimm}, {Kulkarni}, {Kumar},
  {Lebrun}, {Lloyd-Ronning}, {Markwardt}, {Mattson}, {Mushotzky}, {Norris},
  {Osborne}, {Paczynski}, {Palmer}, {Park}, {Parsons}, {Paul}, {Rees},
  {Reynolds}, {Rhoads}, {Sasseen}, {Schaefer}, {Short}, {Smale}, {Smith},
  {Stella}, {Tagliaferri}, {Takahashi}, {Tashiro}, {Townsley}, {Tueller},
  {Turner}, {Vietri}, {Voges}, {Ward}, {Willingale}, {Zerbi}, \&
  {Zhang}}]{gehr04}
{Gehrels}, N., {Chincarini}, G., {Giommi}, P., {et~al.} 2004, Astrophys. J.,
  611, 1005

\bibitem[{{Godet} {et~al.}(2009){Godet}, {Barret}, {Webb}, {Farrell}, \&
  {Gehrels}}]{gode09}
{Godet}, O., {Barret}, D., {Webb}, N.~A., {Farrell}, S.~A., \& {Gehrels}, N.
  2009, Astrophys. J., 705, L109

\bibitem[{{Godet} {et~al.}(2012){Godet}, {Plazolles}, {Kawaguchi}, {Lasota},
  {Barret}, {Farrell}, {Braito}, {Servillat}, {Webb}, \& {Gehrels}}]{gode12}
{Godet}, O., {Plazolles}, B., {Kawaguchi}, T., {et~al.} 2012, \apj, 752, 34

\bibitem[{{Hut} {et~al.}(1992){Hut}, {McMillan}, {Goodman}, {Mateo}, {Phinney},
  {Pryor}, {Richer}, {Verbunt}, \& {Weinberg}}]{hut92}
{Hut}, P., {McMillan}, S., {Goodman}, J., {et~al.} 1992, \pasp, 104, 981

\bibitem[{{Kawaguchi}(2003)}]{kawa03}
{Kawaguchi}, T. 2003, \apj, 593, 69

\bibitem[{{Kawaguchi} {et~al.}(2004){Kawaguchi}, {Aoki}, {Ohta}, \&
  {Collin}}]{kawa04}
{Kawaguchi}, T., {Aoki}, K., {Ohta}, K., \& {Collin}, S. 2004, \aap, 420, L23

\bibitem[{{King}(2008)}]{king08}
{King}, A.~R. 2008, \mnras, 385, L113

\bibitem[{{K{\"o}rding} {et~al.}(2005){K{\"o}rding}, {Colbert}, \&
  {Falcke}}]{koer05}
{K{\"o}rding}, E., {Colbert}, E., \& {Falcke}, H. 2005, Astron. Astrophys.,
  436, 427

\bibitem[{{K{\"o}rding} {et~al.}(2006){K{\"o}rding}, {Falcke}, \&
  {Corbel}}]{koer06}
{K{\"o}rding}, E., {Falcke}, H., \& {Corbel}, S. 2006, Astron. Astrophys., 456,
  439

\bibitem[{{Lasota} {et~al.}(2011){Lasota}, {Alexander}, {Dubus}, {Barret},
  {Farrell}, {Gehrels}, {Godet}, \& {Webb}}]{laso11}
{Lasota}, J.-P., {Alexander}, T., {Dubus}, G., {et~al.} 2011, \apj, 735, 89

\bibitem[{{Madau} \& {Rees}(2001)}]{mada01}
{Madau}, P. \& {Rees}, M.~J. 2001, \apjl, 551, L27

\bibitem[{{Madau} {et~al.}(2004){Madau}, {Rees}, {Volonteri}, {Haardt}, \&
  {Oh}}]{mada04}
{Madau}, P., {Rees}, M.~J., {Volonteri}, M., {Haardt}, F., \& {Oh}, S.~P. 2004,
  \apj, 604, 484

\bibitem[{{Merloni} {et~al.}(2003){Merloni}, {Heinz}, \& {di Matteo}}]{merl03}
{Merloni}, A., {Heinz}, S., \& {di Matteo}, T. 2003, Mon. Not. R. Astron. Soc.,
  345, 1057

\bibitem[{{Miller} \& {Colbert}(2004)}]{mill04}
{Miller}, M.~C. \& {Colbert}, E.~J.~M. 2004, International Journal of Modern
  Physics D, 13, 1

\bibitem[{{Roberts}(2007)}]{robe07}
{Roberts}, T.~P. 2007, \apss, 311, 203

\bibitem[{{Servillat} {et~al.}(2011){Servillat}, {Farrell}, {Lin}, {Godet},
  {Barret}, \& {Webb}}]{serv11}
{Servillat}, M., {Farrell}, S.~A., {Lin}, D., {et~al.} 2011, Astrophys. J.,
  743, 6

\bibitem[{{Soria} {et~al.}(2012){Soria}, {Hakala}, {Hau}, {Gladstone}, \&
  {Kong}}]{sori12}
{Soria}, R., {Hakala}, P.~J., {Hau}, G.~K.~T., {Gladstone}, J.~C., \& {Kong},
  A.~K.~H. 2012, \mnras, 420, 3599

\bibitem[{{Soria} {et~al.}(2010){Soria}, {Hau}, {Graham}, {Kong}, {Kuin}, {Li},
  {Liu}, \& {Wu}}]{sori10}
{Soria}, R., {Hau}, G.~K.~T., {Graham}, A.~W., {et~al.} 2010, Mon. Not. R.
  Astron. Soc., 405, 870

\bibitem[{{Webb} {et~al.}(2012){Webb}, {Cseh}, {Lenc}, {Godet}, {Barret},
  {Corbel}, {Farrell}, {Fender}, {Gehrels}, \& {Heywood}}]{webb12}
{Webb}, N., {Cseh}, D., {Lenc}, E., {et~al.} 2012, Science, 337, 554

\bibitem[{{Webb} {et~al.}(2010){Webb}, {Barret}, {Godet}, {Servillat},
  {Farrell}, \& {Oates}}]{webb10}
{Webb}, N.~A., {Barret}, D., {Godet}, O., {et~al.} 2010, Astrophys. J., 712,
  L107

\bibitem[{{Wiersema} {et~al.}(2010){Wiersema}, {Farrell}, {Webb}, {Servillat},
  {Maccarone}, {Barret}, \& {Godet}}]{wier10}
{Wiersema}, K., {Farrell}, S.~A., {Webb}, N.~A., {et~al.} 2010, Astrophys. J.,
  721, L102

\end{thebibliography}

\end{document}